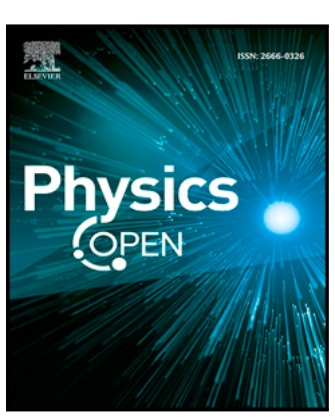

# Machine learning Calabi–Yau three-folds, four-folds, and five-folds

Kaniba Mady Keita [a,b,*], Younouss Hamèye Dicko [b,c]

[a] Centre de Calcul de Modélisation et de Simulation: CCMS Department of Physics, Faculty of Sciences and Techniques, University of Sciences, Techniques and Technologies of Bamako, FST-USTTB, BP:E3206, Mali

[b] Centre de Recherche en Phyique Quantique et de ses Applications: CRPQA, Bamako, Mali

[c] Institut de Consultation et d'Expertise en Education: ICE, Bamako, Mali

ABSTRACT

In this manuscript, we demonstrate, using several regression techniques, that the remaining independent Hodge numbers of complete intersection Calabi–Yau four-folds and five-folds can be machine learned from $h^{1,1}$ and $h^{2,1}$. Consequently, we combine the Hodge numbers $h^{1,1}$ and $h^{2,1}$ from the complete intersection Calabi-Yau three-folds, four-folds, and five-folds into a single dataset. We then implement various classification algorithms on this dataset. For example, Gaussian process and naive Bayes classifiers both achieve 100% accuracy in binary classification between three-folds and four-folds. Using the Support Vector Machine (SVM) algorithm, a special corner is identified in the Calabi–Yau four-fold landscape (characterized by $15 \leq h^{1,1} \leq 30$ and $95 \leq h^{2,1} \leq 100$) during multiclass classification. Furthermore, the highest accuracy 1.00000, in classifying Calabi–Yau three-folds, four-folds, and five-folds is obtained using the naive Bayes classifier.

## 1. Introduction

Conformally Calabi–Yau manifolds have played a significant role in string theory compactification [1–4]. Notably, compactifying various string theories on Calabi–Yau four-folds gives rise to nontrivial superconformal field theories [5]. In the context of M-theory, compactification on mirror Calabi–Yau four-folds leads to the exchange of one of the Higgs branches with the Coulomb branch via mirror symmetry [6]. Additionally, it has been shown that, upon compactification on Calabi–Yau five-folds, the Kähler and complex structure moduli reside in different supermultiplets [7].

These and other results obtained through string-theoretic probes have strongly motivated the systematic construction and classification of Calabi–Yau manifolds [8–11]. In particular, the number of Calabi–Yau manifolds that are complete intersections in products of projective spaces has been computed. Such manifolds are commonly referred to as *complete intersection Calabi–Yau* manifolds, abbreviated as **CICY**. When the complex dimension is specified, we denote them as $\mathbf{CICY}_n$ or **CICY**n, where $n$ represents the complex dimension.

There are 7890 Calabi–Yau three-folds, and the complete list of these manifolds is available in [12]. This number reduces drastically to 266 when considering only distinct Hodge number combinations. Similarly, the number of Calabi–Yau four-folds that are complete intersections is 921,497, as reported in [13], with 4418 distinct Hodge number combinations. For Calabi–Yau five-folds, we consider the 27,068 manifolds constructed in [14], for which cohomological data has been successfully computed for 12,433 of them (53.7%). The resulting dataset contains 2375 distinct Hodge diamonds. In what follows, we use only these 2375 distinct Hodge number combinations from [14]. Therefore, for our classification tasks, we exclusively consider these reduced sets of distinct Hodge numbers: 266 for $\mathrm{CICY}_3$, 4418 for $\mathrm{CICY}_4$, and 2375 for $\mathrm{CICY}_5$. These are then combined into a single dataset to which we apply various machine learning classification techniques. The source files and the combined dataset are available from [15].

The application of machine learning (ML) to Calabi–Yau manifolds has recently surged in popularity [16–31] and it continues to garner significant attention nowadays. These studies have demonstrated that various aspects of CICY manifolds are machine-learnable, particularly through supervised learning techniques such as classification and regression algorithms. The performance metrics of these models often show near-perfect accuracy. However, one should emphasize that these applications encounter a fundamental challenge in ML: the *one-to-multivalue correspondence* problem (i.e., non-unique mappings), as discussed in [32–35]. This challenge becomes even more severe when configuration matrices are used as input features, as is the case in many of the cited works, because the same physical state may correspond to multiple equivalent representations. The higher the degeneracy of the input–output map, the more difficult the learning task becomes,

* Corresponding author at: Centre de Calcul de Modélisation et de Simulation: CCMS Department of Physics, Faculty of Sciences and Techniques, University of Sciences, Techniques and Technologies of Bamako, FST-USTTB, BP:E3206, Mali.

*E-mail address:* kanibamady.keita@usttb.edu.ml (K.M. Keita).

unless the degeneracies are explicitly incorporated into the model architecture-for example, through symmetry-aware [36,37] or equivariant network designs [38,39]. Addressing these degeneracies is therefore crucial for improving model accuracy and stability.

To alleviate this issue, we instead use the Hodge numbers $h^{1,1}$ and $h^{2,1}$ as input features. While the mapping remains many-to-one, this representation simplifies the learning task. Despite this challenge, machine learning can still be effectively applied to explore the CICY landscape. To evaluate and compare the performance of different ML models, we use standard metrics. One such metric is the $R^2$ score in regression tasks, which measures how well the predicted values match the target values. For classification tasks, we use *accuracy*, which is defined as the ratio of correctly predicted labels to the total number of labels in the validation dataset.[1]

In a previous article [40], we applied various regression algorithms to learn the Hodge number $h^{2,1}$ from $h^{1,1}$ for CICY$_3$. Here, we extend this approach to learn $h^{3,1}$ in terms of $h^{1,1}$ and $h^{2,1}$ for CICY$_4$. The performance of these supervised learning models suggests that regression techniques can indeed reduce the number of independent Hodge numbers required to describe a given CICY.

Motivated by this, we compile the Hodge numbers $h^{1,1}$ and $h^{2,1}$ from CICY$_3$, CICY$_4$, and CICY$_5$ into a single dataset and perform classification tasks. The classification algorithms used include: Gaussian Process Regression (gausspr) [41], Naive Bayes [42,43], Support Vector Classifier (SVM) [44–46], Random Forest (RF) [47,48], Logistic Regression [49], Linear Discriminant Analysis (LDA) [50], Recursive Partitioning (rpart) [51], and $k$-Nearest Neighbor (KNN) [52]. Our results show that these models perform well in classifying the CICY manifolds by dimension.

The remainder of this manuscript is organized as follows:

- Section 2 presents an exploratory data analysis (EDA) of the Hodge numbers for CICY$_3$, CICY$_4$, and CICY$_5$. In particular, we analyze the *skewness* of their distributions [53], which is relevant due to its connection with class imbalance in deep learning [54].
- Section 3 applies regression algorithms to predict $h^{3,1}$ from $h^{1,1}$ and $h^{2,1}$ for CICY$_4$, extending the methodology of our earlier work on CICY$_3$ [40].
- Section 4 focuses on classification tasks. We first classify CICY$_3$ versus CICY$_4$, then perform multi-class classification to distinguish between CICY$_3$, CICY$_4$, and CICY$_5$, using the previously listed algorithms.
- Section 4.2 concludes the paper with a summary of our findings and discusses possible directions for future work.

## 2. Skewness analysis of CICY three-, four-, and five-folds

In this section, we perform a skewness analysis of the datasets corresponding to CICY$_3$, CICY$_4$, and CICY$_5$. While the skewness of the dataset is expected not to alter the precision of machine learning classifications [55], its connection to the class imbalance problem in deep learning compels us to investigate it further.

We begin with a brief overview of some fundamental properties of CICYs.

Calabi–Yau manifolds $\mathbb{X}$ of complex dimension $n$, considered as complete intersections in products of complex projective spaces, are characterized by a *configuration matrix* $\mathbb{C}$, defined as [11]:

$$\mathbb{C} = \left[\begin{array}{c||cccc} \mathbb{P}^{n_1} & q_1^1 & \cdots & q_K^1 \\ \mathbb{P}^{n_2} & q_1^2 & \cdots & q_K^2 \\ \vdots & \vdots & \ddots & \vdots \\ \mathbb{P}^{n_m} & q_1^m & \cdots & q_K^m \end{array}\right] = \left[\begin{array}{c||cccc} n_1 & q_1^1 & \cdots & q_K^1 \\ n_2 & q_1^2 & \cdots & q_K^2 \\ \vdots & \vdots & \ddots & \vdots \\ n_m & q_1^m & \cdots & q_K^m \end{array}\right], \quad q_a^r \in \mathbb{Z}_{\geq 0}. \tag{2.1}$$

[1] Simple explanations of accuracy and other statistical measures are available in [27,28].

The manifold $\mathbb{X}$ is then a complete intersection of $K$ transverse hypersurfaces (homogeneous holomorphic polynomials) $\mathbb{X}_1, \ldots, \mathbb{X}_K$ in a product of projective spaces $\mathbb{P}^n$, where the degrees of these hypersurfaces with respect to $\mathbb{P}^{n_a}$ are given by $q_1^a, q_2^a, \ldots, q_K^a$. These degrees and other parameters in the configuration matrix $\mathbb{C}$ must satisfy the following conditions:

$$\sum_{r=1}^{m} n_r = n + K, \qquad \sum_{\alpha=1}^{K} q_\alpha^r = n_r + 1, \quad \text{for all } r = 1, \ldots, m. \tag{2.2}$$

For each CICY, one can compute the dimensions of the Dolbeault cohomology groups $H^{p,q}$, known as the Hodge numbers $h^{p,q}$. These satisfy Poincaré duality ($h^{p,q} = h^{n-p,n-q}$) and holomorphic/antiholomorphic symmetry ($h^{p,q} = h^{q,p}$) [56]. The Hodge numbers count the number of harmonic $(p,q)$-forms on $\mathbb{X}$. The Euler characteristic is given in terms of the Hodge numbers as:

$$\chi = \sum_{p,q} (-1)^{p+q} h^{p,q}. \tag{2.3}$$

For any CICY, the top holomorphic form $(n,0)$ always satisfies $h^{n,0} = 1$, and if the manifold is simply connected (as is the case for most CICYs), then $h^{1,0} = 0$. These restrictions determine fixed values at the corners of the so-called Hodge diamond. However, the remaining independent Hodge numbers can still exhibit skewness.

Skewness measures the asymmetry of a distribution and is closely related to the class imbalance problem in machine learning. The generic formula for computing the skewness of a given Hodge number $h^{p,q}$ across a dataset of CICYs is:

$$\text{Skewness} = \frac{\sum_{k=1}^{N} \left(C_k^{p,q} - \bar{C}^{p,q}\right)^3}{(N-1)\sigma^3}, \tag{2.4}$$

where $\bar{C}^{p,q}$ is the mean value of the column for $h^{p,q}$, $C_k^{p,q}$ is the value of the $k$th CICY for the same Hodge number, $N$ is the number of CICYs in the dataset, and $\sigma$ denotes the standard deviation of the values in that column.

In the following plots and throughout the analysis, the Hodge numbers $h^{p,q}$ will be abbreviated as $Hpq$.

### 2.1. Skewness of the Hodge numbers for CICY$_3$

Based on the general properties of CICYs discussed above, we note that CICY$_3$ has only two independent Hodge numbers: $h^{1,1}$ and $h^{2,1}$. The Euler characteristic in this case simplifies to:

$$\chi = 2(h^{1,1} - h^{2,1}). \tag{2.5}$$

The skewness of these Hodge numbers is analyzed and visualized in Fig. 1.

The plot of $h^{2,1}$ as a function of $h^{1,1}$ is shown in Fig. 2.

### 2.2. Skewness of the Hodge numbers for CICY$_4$

The Hodge diamond of CICY$_4$ has three independent Hodge numbers: $h^{1,1}$, $h^{1,2}$, and $h^{1,3}$. This reduction in the number of independent quantities is due to the existence of a notable linear relation among the Hodge numbers [57]:

$$h^{2,2} = 2\left(22 + 2h^{1,3} + 2h^{1,1} - h^{1,2}\right). \tag{2.6}$$

As a result, in Section 3, we apply various regression techniques to machine-learn $h^{1,3}$ in terms of $h^{1,1}$ and $h^{1,2}$. The skewness of the Hodge numbers for CICY$_4$ is shown in Fig. 3.

In addition, plots of the other Hodge numbers as functions of $h^{1,1}$ and $h^{2,1}$ for CICY$_4$ are presented. These visualizations underscore the inherent challenges in applying machine learning algorithms to the CICY$_4$ dataset (see Figs. 4 and 5).

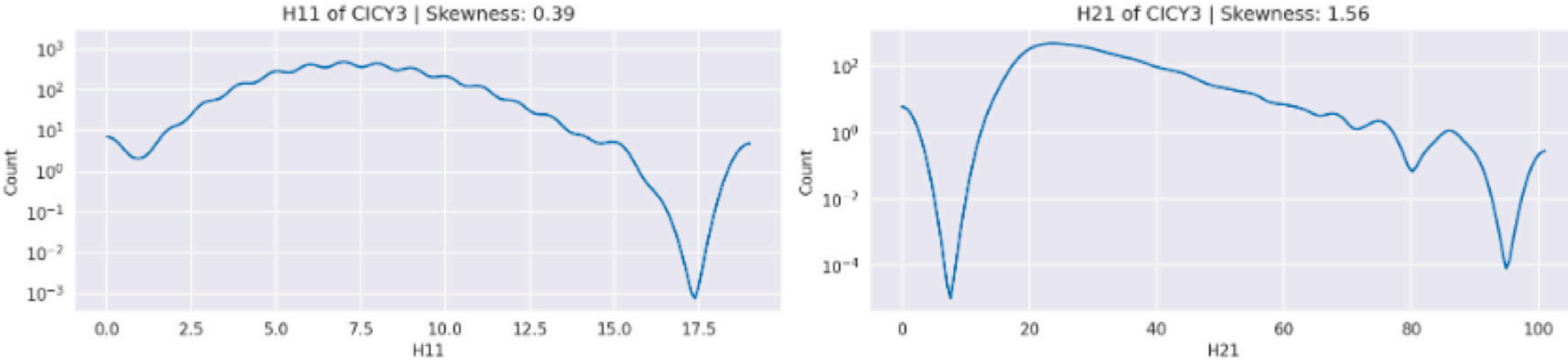

**Fig. 1.** Skewness of the Hodge numbers for CICY3 with a log-scaled count axis.

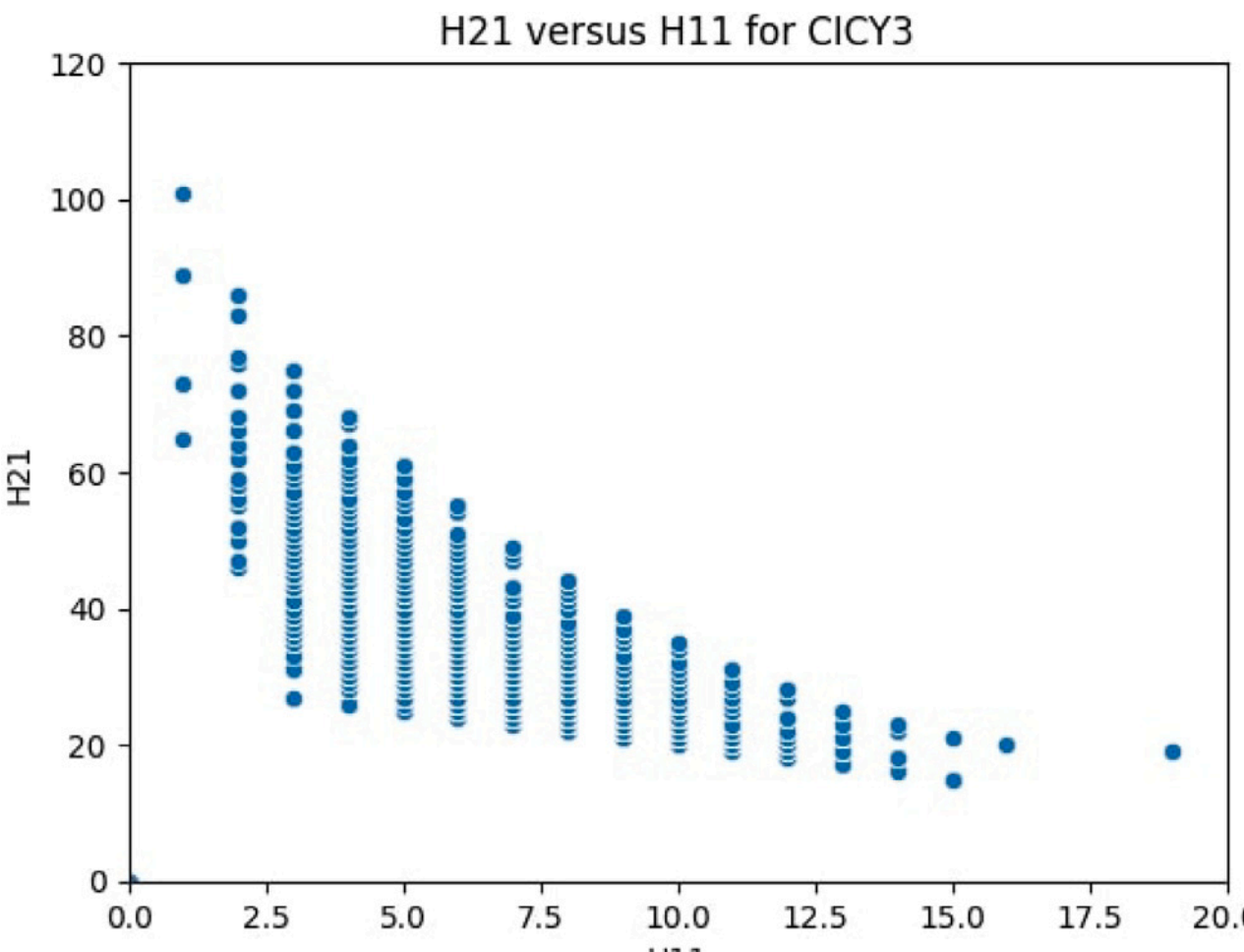

**Fig. 2.** $h^{2,1}$ as function of $h^{1,1}$ for CICY3.

### 2.3. Skewness of the Hodge numbers for CICY$_5$

The total number of CICY$_5$ manifolds is not yet fully determined. However, the dataset presented in Ref. [14] provides a valuable foundation for analysis.

For CICY$_5$, the Calabi–Yau condition imposes $h^{0,0} = h^{5,5} = 1$ and $h^{1,0} = h^{0,1} = h^{4,5} = h^{5,4} = 0$. Due to the symmetries and dualities of the Hodge numbers, the independent (unspecified) Hodge numbers reduce to:

$$h^{1,1}, \quad h^{2,1}, \quad h^{3,1}, \quad h^{4,1}, \quad h^{2,2}, \quad \text{and} \quad h^{3,2}.$$

The skewness of these Hodge numbers, based on the dataset from Ref. [14], is presented in Fig. 6.

**Interpretation of Skewness Analysis:** Throughout all these graphs, the skewness analysis reveals systematic asymmetries in the distributions of independent Hodge numbers across CICY$_3$, CICY$_4$, and CICY$_5$ datasets. In CICY$_3$, $h^{1,1}$ and $h^{2,1}$ are moderately skewed, indicating relatively balanced classes and easier learning for ML models. In contrast, several Hodge numbers in CICY$_4$ and CICY$_5$ show more pronounced skewness, reflecting underrepresented classes that can challenge regression or classification tasks. It should stress that such imbalances motivate strategies like data augmentation, weighting, or symmetry-aware models, while the analysis overall provides a quantitative characterization of dataset structure and guides model design and feature selection.

## 3. Learning $\hat{h}^{3,1} = \mathcal{F}\left(h^{1,1}, h^{2,1}\right)$ for the reduced CICY$_4$ dataset

Before proceeding to the classification tasks, we aim to machine-learn the Hodge number $h^{3,1}$ using $h^{1,1}$ and $h^{2,1}$ as input features. Specifically, we seek to approximate a function $\mathcal{F}$ that captures the relationship:

$$\hat{h}^{3,1} = \mathcal{F}\left(h^{1,1}, h^{2,1}\right),$$

while keeping in mind that this mapping is inherently many-to-one.

To this end, we train various regression models using 80% of the reduced CICY$_4$ dataset, which contains 4418 entries. The remaining 20% of the data is reserved for validating the trained models.

The performance of each regression model is evaluated using two standard metrics: the **root mean square error (RMSE)** and the **Pearson correlation coefficient** ($R^2$). All algorithms are implemented in `RStudio`, and the hyperparameters used match those employed in our earlier study on CICY$_3$ (see Ref. [40]).

The results obtained from these regression models are summarized in Table 1.

These results indicate that the Hodge number $h^{3,1}$ in the reduced dataset of 4418 CICY$_4$ manifolds is effectively learnable from $h^{1,1}$ and $h^{2,1}$ using machine learning techniques. In particular, Gaussian Process Regression (GPR) demonstrates excellent performance on this dataset. A plot comparing the predicted values from GPR with the true values is shown in Fig. 7.

The poorest performance is observed with Extreme Gradient Boosting (XGBoost), and the corresponding plots are shown in Fig. 8.

As for the CICY$_5$ dataset, a similar investigation can be conducted to learn the remaining Hodge numbers in terms of $h^{1,1}$ and $h^{2,1}$ for the 2375 CICY$_5$ manifolds.

For instance, learning $h^{3,1}$ from $h^{1,1}$ and $h^{2,1}$ using Gaussian Process Regression yields excellent results:

$$\text{Validation set: } R^2 = 100\%, \quad \text{RMSE} = 0.0005671492,$$

$$\text{Training set: } R^2 = 100\%, \quad \text{RMSE} = 0.0005492653.$$

In contrast, Extreme Gradient Boosting performs poorly on this task, with

$$\text{Validation set: } R^2 = 20.26\%, \quad \text{RMSE} = 7.148835.$$

These findings indicate that an interesting and learnable combined dataset can be constructed using CICY$_3$, CICY$_4$, and CICY$_5$ manifolds.

Consequently, we now consider $h^{1,1}$ and $h^{2,1}$ from CICY$_3$, CICY$_4$, and CICY$_5$ as features for our classification tasks.

## 4. Classification of CICY manifolds

The goal of this section is to classify CICY$_3$, CICY$_4$, and CICY$_5$ manifolds using only the Hodge numbers $h^{1,1}$ and $h^{2,1}$. The performance of the classification models is evaluated using the so-called *Confusion Matrix*. For classification involving $N$ classes, this matrix has the generic form:

$$\text{Confusion Matrix} = \begin{bmatrix} C_{11} & C_{12} & \cdots & C_{1N} \\ C_{21} & C_{22} & \cdots & C_{2N} \\ \vdots & \vdots & \ddots & \vdots \\ C_{N1} & C_{N2} & \cdots & C_{NN} \end{bmatrix}$$

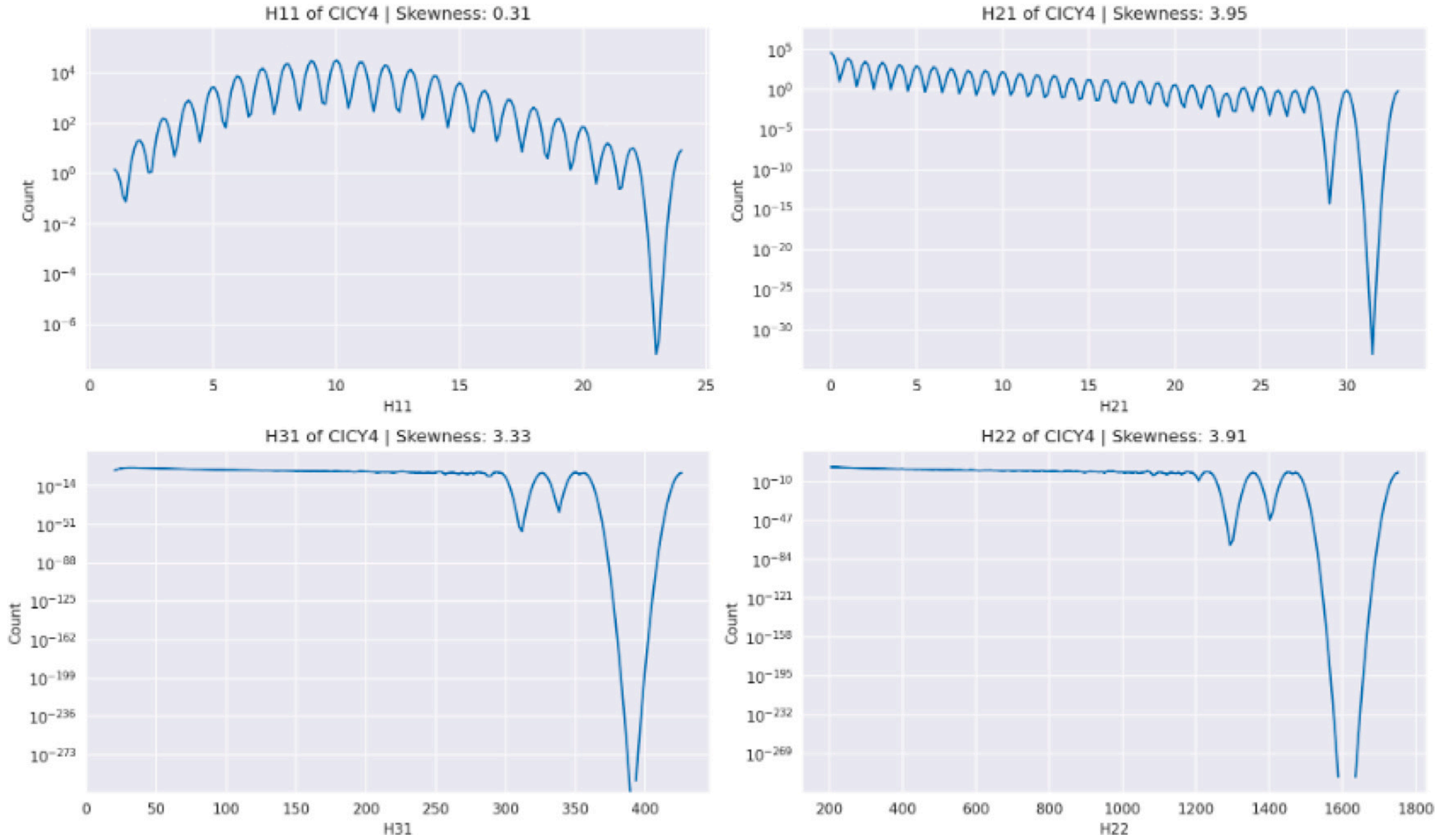

**Fig. 3.** Skewness of the Hodge numbers for CICY4 with a log-scaled count axis.

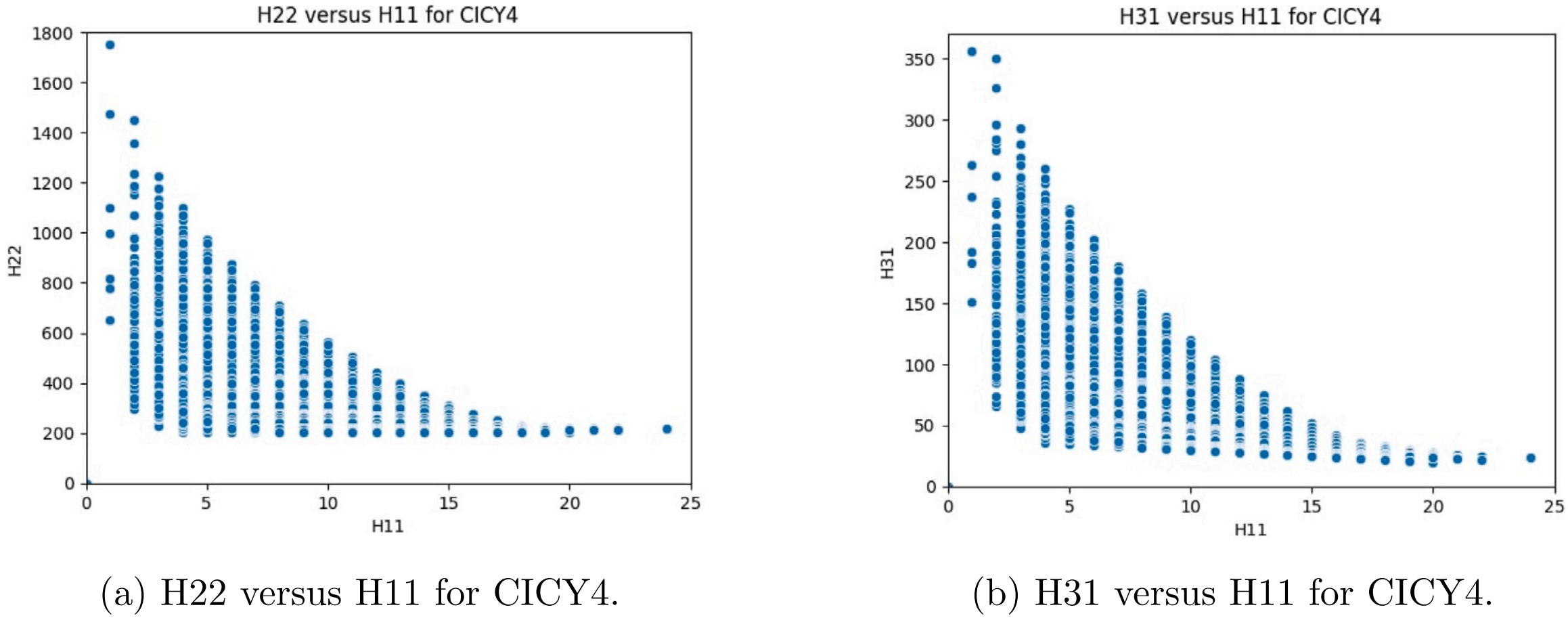

(a) H22 versus H11 for CICY4.

(b) H31 versus H11 for CICY4.

**Fig. 4.** The plot of the other Hodge numbers versus H11 for CICY4.

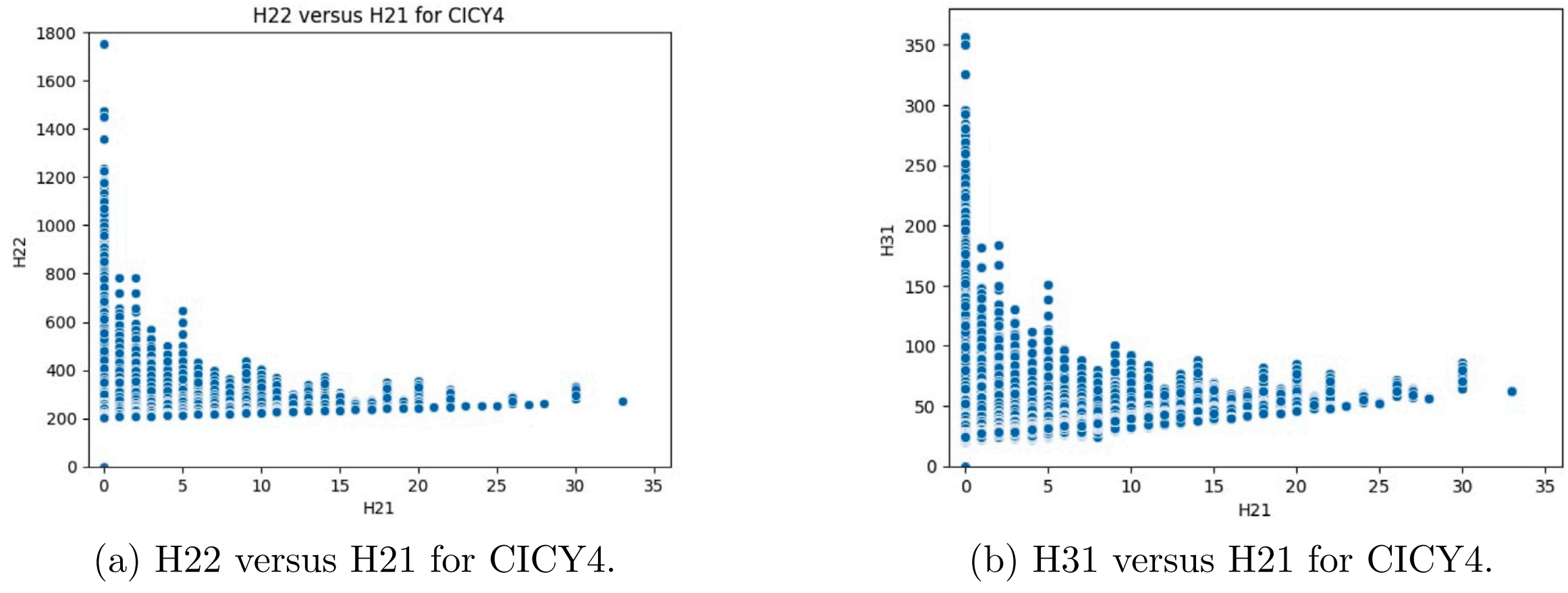

(a) H22 versus H21 for CICY4.

(b) H31 versus H21 for CICY4.

**Fig. 5.** The plot of the other Hodge numbers versus H21 for CICY4.

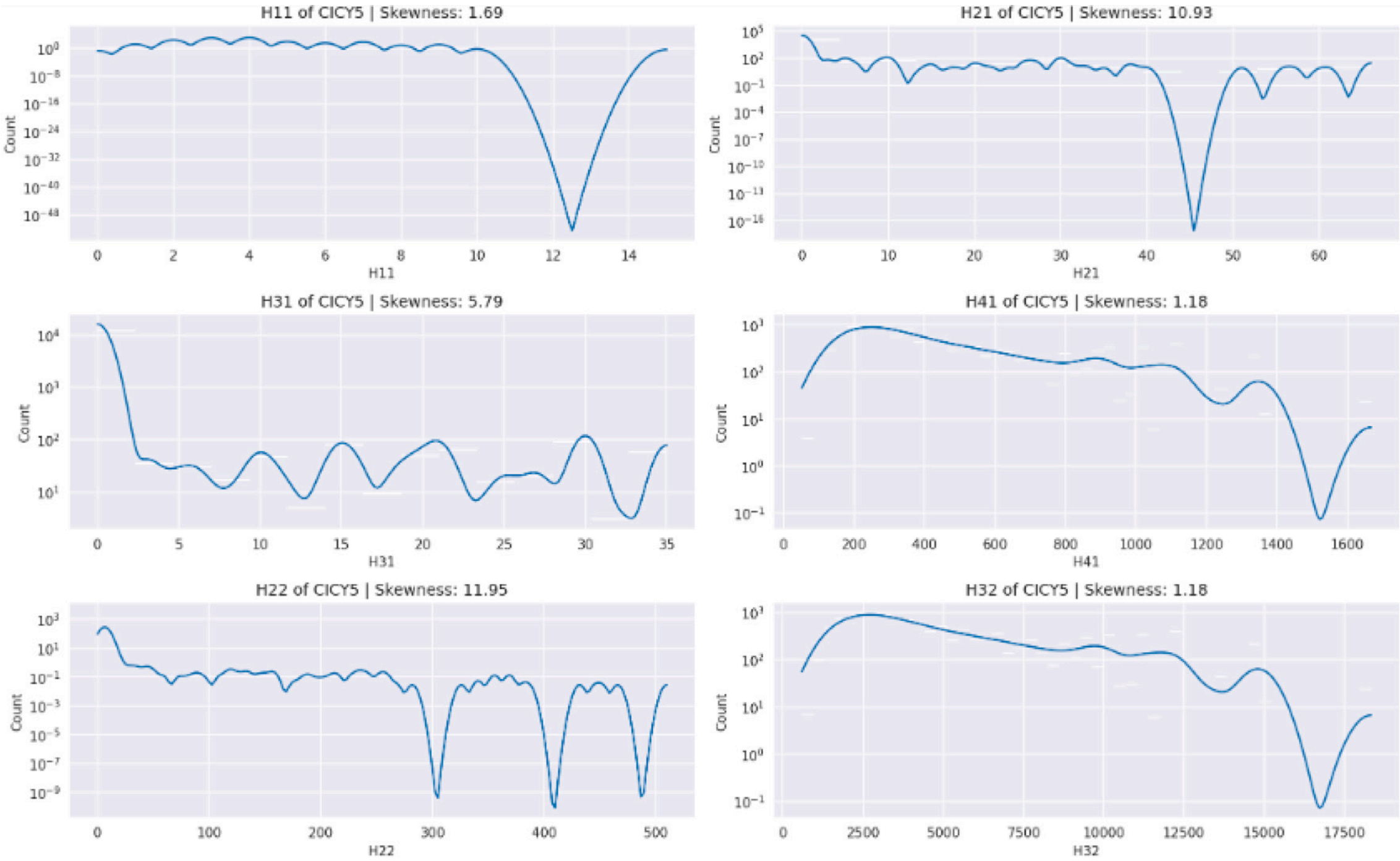

**Fig. 6.** Skewness of the Hodge numbers for CICY5 with a log-scaled count axis.

**Table 1**
Statistical measures for models applied to the 4418 of CICY4.

| Regression | Validation | | | Calibration | | |
|---|---|---|---|---|---|---|
| | $R^2$ | RMSE | P-value | $R^2$ | RMSE | P-value |
| gausspr | 0.999999999 | 0.00207779 | $<2.2 \times 10^{-16}$ | 0.999999999 | 0.002105201 | $<2.2 \times 10^{-16}$ |
| KSVM | 0.9990968 | 2.510213 | $<2.2 \times 10^{-16}$ | 0.9990771 | 2.466058 | $<2.2 \times 10^{-16}$ |
| RF | 0.9969507 | 2.190028 | $<2.2 \times 10^{-16}$ | 0.9986549 | 1.555032 | $<2.2 \times 10^{-16}$ |
| Xgboost | 0.6094061 | 27.90818 | $<2.2 \times 10^{-16}$ | 0.5917905 | 25.56546 | $<2.2 \times 10^{-16}$ |

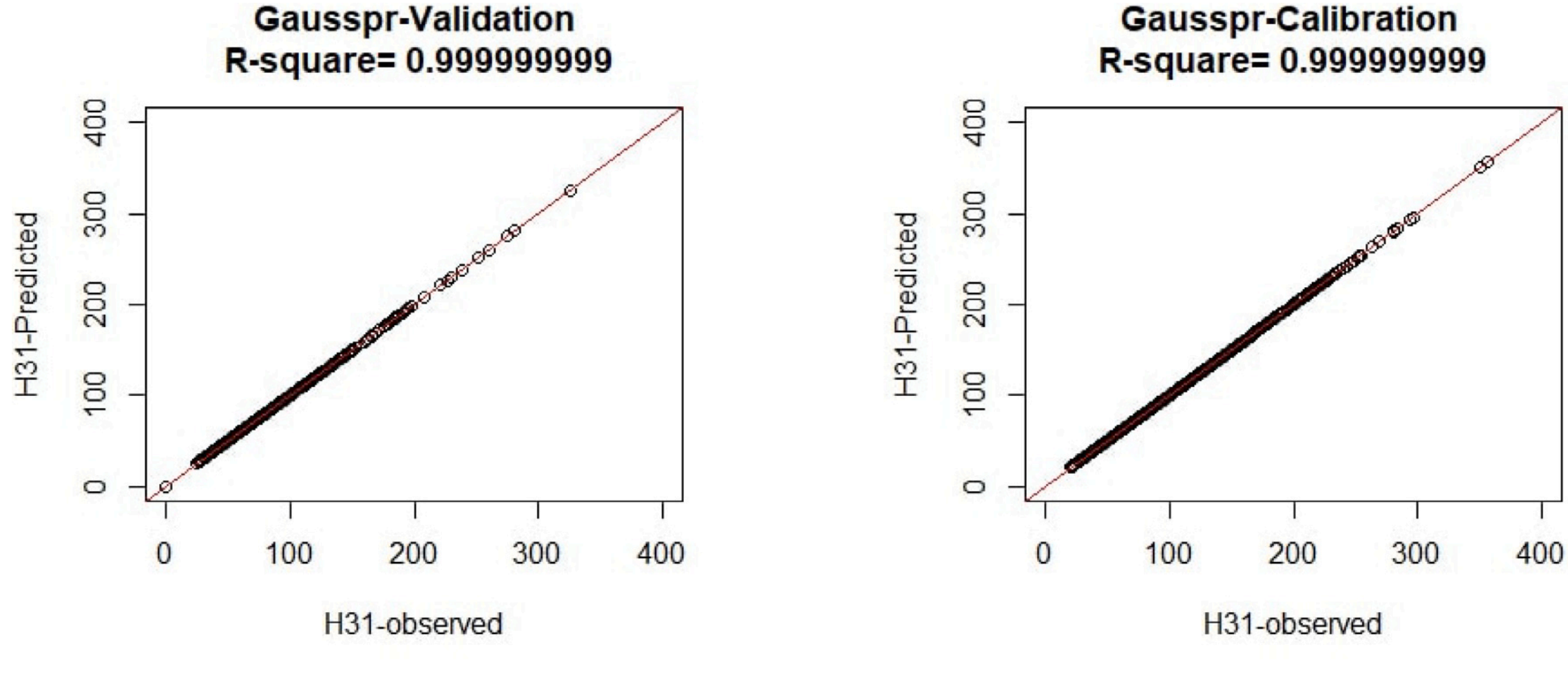

(a) gausspr for the validation set. (b) Gausspr for the Calibration set.

**Fig. 7.** The predictions from Gausspr for the reduced 4418 of CICY4.

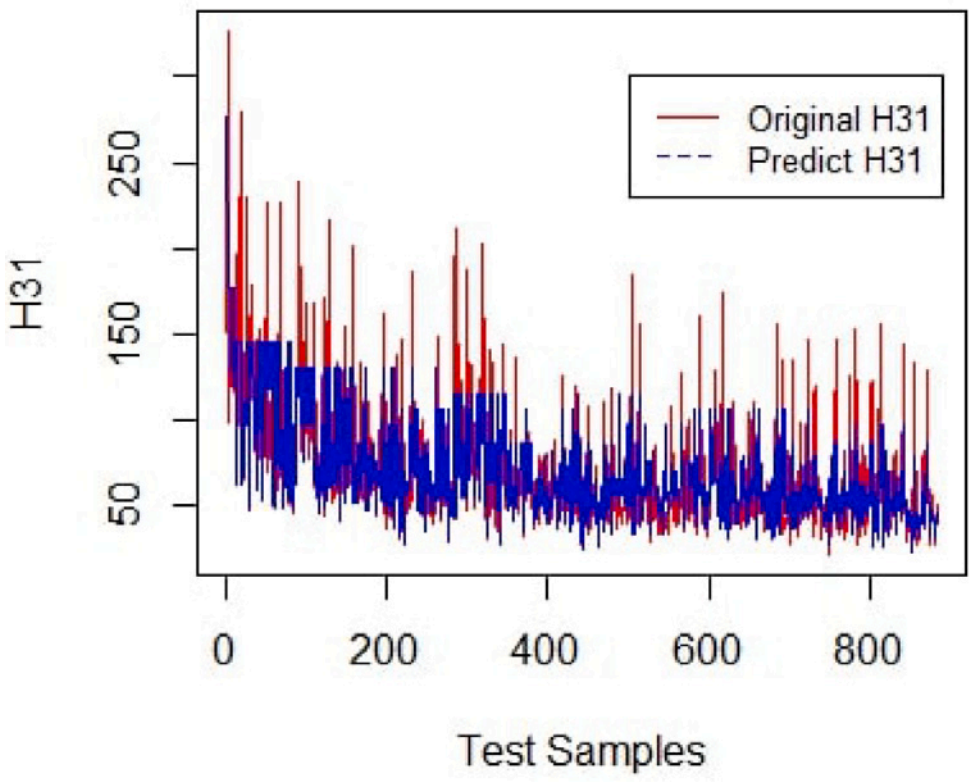

(a) Extreme gradient boost for CICY4.

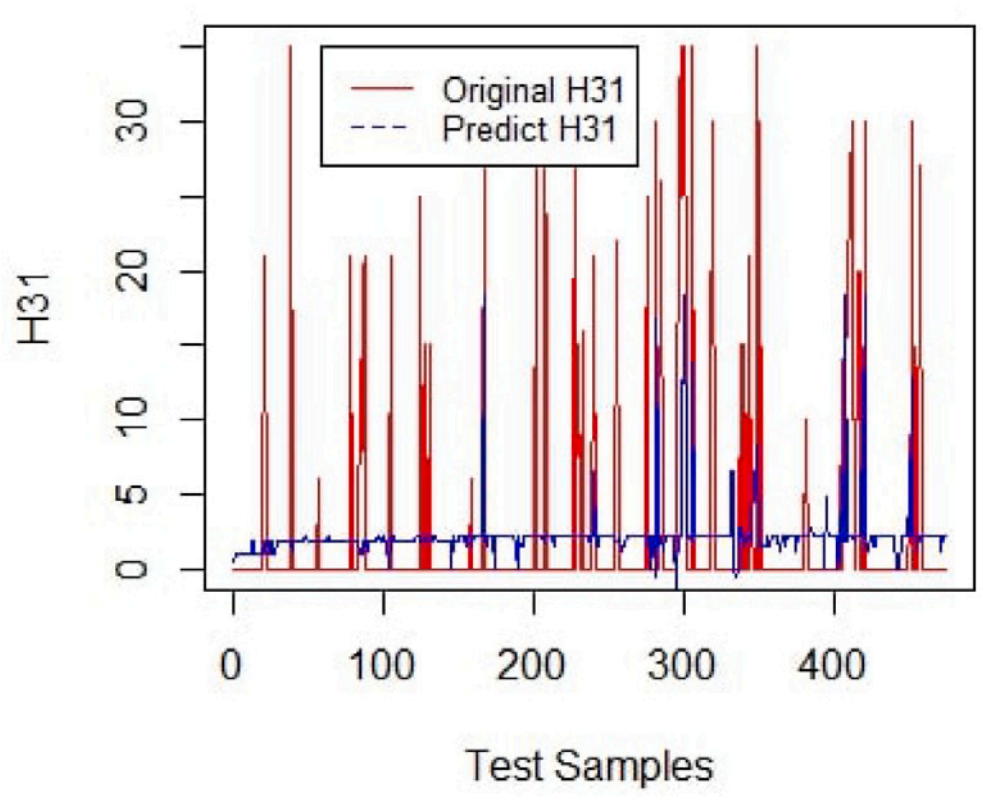

(b) Extreme gradient boost for CICY5.

**Fig. 8.** The plot of the Extreme gradient boost for CICY4 and CICY5.

Alternatively, this can be labeled as follows:

Confusion Matrix

$$= \begin{bmatrix} \text{True}\backslash\text{Predicted} & \text{Class 1} & \text{Class 2} & \cdots & \text{Class } N \\ \text{Class 1} & C_{11} & C_{12} & \cdots & C_{1N} \\ \text{Class 2} & C_{21} & C_{22} & \cdots & C_{2N} \\ \vdots & \vdots & \vdots & \ddots & \vdots \\ \text{Class } N & C_{N1} & C_{N2} & \cdots & C_{NN} \end{bmatrix}$$

Here, $C_{ii}$ represents the number of correct predictions for class $i$, whereas $C_{ij}$ $(i \neq j)$ denotes the number of instances from class $i$ misclassified as class $j$.

To evaluate the classification performance, we consider the **accuracy** and the **F1-score**. For binary classification, a single F1-score is computed, while for multiclass classification, one F1-score is computed per class. These metrics are derived from the confusion matrix [27,58,59].

In order to compute accuracy and F1-score for class $i$, the following quantities are defined [27]:

- **True Positive (TP)**: Correctly predicted positives (diagonal entry).
- **False Positive (FP)**: Instances incorrectly predicted as class $i$ (column-wise, excluding the diagonal).
- **False Negative (FN)**: Instances of class $i$ misclassified as another class (row-wise, excluding the diagonal).
- **True Negative (TN)**: All other correctly predicted instances (excluding the row and column for class $i$).

Mathematically, these are defined as:

$$\begin{aligned} \text{TP} &= C_{ii} \\ \text{FP} &= \sum_{j \neq i} C_{ji} \\ \text{FN} &= \sum_{j \neq i} C_{ij} \\ \text{TN} &= \sum_{\substack{j=1 \\ j \neq i}}^{N} C_{jj} \end{aligned} \tag{4.1}$$

The **accuracy** is defined as the total number of correct predictions divided by the total number of samples:

$$\begin{aligned} \text{Accuracy} &= \frac{\sum_{i=1}^{N} C_{ii}}{\sum_{i=1}^{N} \sum_{j=1}^{N} C_{ij}} \\ &= \frac{\text{Number of Correct Predictions}}{\text{Total Number of Predictions}} \\ &= \frac{\text{TP} + \text{TN}}{\text{TP} + \text{TN} + \text{FP} + \text{FN}}. \end{aligned} \tag{4.2}$$

The **F1-score** for class $i$ is computed using precision and recall:

$$\begin{aligned} \text{Precision} &= \frac{\text{TP}}{\text{TP} + \text{FP}} = \frac{C_{ii}}{C_{ii} + \sum_{j \neq i} C_{ji}} \\ \text{Recall} &= \frac{\text{TP}}{\text{TP} + \text{FN}} = \frac{C_{ii}}{C_{ii} + \sum_{j \neq i} C_{ij}} \\ \text{F1-score} &= \frac{2 \times \text{Precision} \times \text{Recall}}{\text{Precision} + \text{Recall}} \end{aligned} \tag{4.3}$$

This formalism is essential for interpreting the confusion matrix in our classification context. To ensure the integrity of the results, we eliminate any overlapping data points i.e., those $\text{CICY}_3$ and $\text{CICY}_4$ manifolds with identical values for both $h^{1,1}$ and $h^{2,1}$ prior to performing the classification.

We are now in a position to classify CICYs using only $h^{1,1}$ and $h^{2,1}$ as input features for our machine learning algorithms. We begin with a binary classification task distinguishing between $\text{CICY}_3$ and $\text{CICY}_4$ manifolds. All algorithms are implemented in R using standard libraries. The code up to data splitting is provided in [15], and the remaining code is available upon request.

### 4.1. Binary classification of $\text{CICY}_3$ and $\text{CICY}_4$

Binary classification serves as a preparatory step toward multiclass classification. The full dataset consists of 266 $\text{CICY}_3$ and 4418 $\text{CICY}_4$ manifolds. We observe that 102 data points are shared between both datasets (i.e., they have identical $h^{1,1}$ and $h^{2,1}$ values) and are thus removed. This results in a reduced dataset containing 216 $\text{CICY}_3$ and 4366 $\text{CICY}_4$ entries.

This dataset presents a class imbalance issue due to the relatively small number of $\text{CICY}_3$ instances. To address this, we apply a combination of the SMOTE and Tomek Links algorithms [60] to balance the dataset. The resulting balanced dataset contains 2365 $\text{CICY}_3$ and 4366 $\text{CICY}_4$ manifolds. This dataset is then split into a training set (80%) and a validation set (20%).

The resulting confusion matrices and performance metrics (accuracy and F1-score) for the validation and training sets are presented in Tables 2 and 3, respectively. Notably, the Gaussian Process and Naive Bayes classifiers show excellent performance in distinguishing $\text{CICY}_3$ from $\text{CICY}_4$.

**Table 2**
Binary classifications of CICY3& 4 validation sets.

| Classification | Accuracy | F1-score | Confusion matrix | | |
|---|---|---|---|---|---|
| Gausspr | 1.00000000 | 1.00000 | Class | CICY3 | CICY4 |
| | | | CICY3 | **501** | 0 |
| | | | CICY4 | 0 | **846** |
| Naive Bayes | 1.0000000 | 1.00000 | Class | CICY3 | CICY4 |
| | | | CICY3 | **501** | 0 |
| | | | CICY4 | 0 | **846** |
| Knn | 0.9970304 | 0.9960 | Class | CICY3 | CICY4 |
| | | | CICY3 | **499** | 2 |
| | | | CICY4 | 2 | **844** |
| svm | 0.9970304 | 0.9960 | Class | CICY3 | CICY4 |
| | | | CICY3 | **499** | 2 |
| | | | CICY4 | 2 | **844** |
| Rpart | 0.9948033 | 0.9930 | Class | CICY3 | CICY4 |
| | | | CICY3 | **499** | 2 |
| | | | CICY4 | 5 | **841** |
| Logistic | 0.9970304 | 0.9976 | Class | CICY3 | CICY4 |
| | | | CICY3 | **499** | 2 |
| | | | CICY4 | 2 | **844** |
| Lda | 0.9873794 | 0.9827 | Class | CICY3 | CICY4 |
| | | | CICY3 | **484** | 17 |
| | | | CICY4 | 0 | **846** |

**Table 3**
Binary classifications of CICY3&4 training sets.

| Classification | Accuracy | F1-score | Confusion matrix | | |
|---|---|---|---|---|---|
| Gausspr | 1.00000000 | 1.00000 | Class | CICY3 | CICY4 |
| | | | CICY3 | **1864** | 0 |
| | | | CICY4 | 0 | **3520** |
| Naive Bayes | 0.9990000 | 0.9992 | Class | CICY3 | CICY4 |
| | | | CICY3 | **11864** | 0 |
| | | | CICY4 | 3 | **3517** |
| Knn | 0.9973997 | 0.9963 | Class | CICY3 | CICY4 |
| | | | CICY3 | **1864** | 0 |
| | | | CICY4 | 14 | **3506** |
| Rpart | 0.9955423 | 0.9936 | Class | CICY3 | CICY4 |
| | | | CICY3 | **1856** | 8 |
| | | | CICY4 | 16 | **3504** |
| svm | 0.9968425 | 0.9955 | Class | CICY3 | CICY4 |
| | | | CICY3 | **1864** | 60 |
| | | | CICY4 | 17 | **3505** |
| Logistic | 0.9977712 | 0.9983 | Class | CICY3 | CICY4 |
| | | | CICY3 | **1864** | 0 |
| | | | CICY4 | 12 | **3508** |
| Lda | 0.9888559 | 0.9837 | Class | CICY3 | CICY4 |
| | | | CICY3 | **1811** | 53 |
| | | | CICY4 | 7 | **3513** |

These tables provide a compact summary of classification performance. The diagonal elements of each confusion matrix correspond to correctly classified instances, while off-diagonal elements represent misclassified entries. For example, in the validation set, the logistic regression model misclassifies **two CICY$_3$** instances as CICY$_4$ and **one CICY$_4$** as CICY$_3$.

This binary classification is important not only as a preliminary task but also in identifying naturally strong binary classifiers for CICY datasets—Support Vector Machines (SVMs), in particular, are well-known for this role [28].

In the subsequent multiclass classification task, we omit Gaussian Process classification and focus on Naive Bayes and other established classifiers.

These results demonstrate that CICY manifolds can be effectively classified using only the two Hodge numbers $h^{1,1}$ and $h^{2,1}$ as input features. Among the classification models tested, the Gaussian Process and Naive Bayes classifiers exhibit the best overall performance.

Furthermore, Fig. 9 displays the Naive Bayes density plots for the two Hodge numbers associated with CICY$_3$ and CICY$_4$. These plots represent the estimated likelihood distributions of the Hodge numbers within each class, providing insight into how the model distinguishes between the two types of CICY manifolds.

In the Naive Bayes density plots, there is no overlap between the distributions of CICY$_3$ and CICY$_4$. This clear separation indicates that the two Hodge numbers, $h^{1,1}$ and $h^{2,1}$, serve as highly discriminative features for distinguishing between these two classes. The well-separated densities confirm the effectiveness of these invariants in classification.

On the other hand, the decision boundary graph for the Support Vector Machine (Fig. 10(a)) reveals some overlap, with portions of CICY$_3$ intersecting the region of CICY$_4$. This suggests that the SVM classifier faces more difficulty in perfectly separating the two classes based on these features.

In these graphs, the decision boundary is clearly visible, and the manifolds are well separated. We observe a separable classification using a linear kernel, and the RBF kernel yields essentially the same results in our model.

### 4.2. Multiclass classification of CICY3, CICY4, and CICY5

Building on the successful binary classification discussed above, we now apply the algorithms to classify CICY datasets using only the Hodge numbers $h^{1,1}$ and $h^{2,1}$. In the constructed dataset, we identified 4012 observables that belong to more than one class; consequently, these were removed. The resulting dataset contains 3047 unique observables, including only 4 belonging to the CICY5 class. This creates a significant class imbalance problem, particularly for CICY5 [61,62], which is a common challenge in machine learning.

To address this, we adopt an oversampling strategy [63,64] to balance the dataset.[2] After oversampling, the dataset consists of 8514 observables, and the classification algorithms are applied to this balanced set.

Due to the limited number of CICY5 samples, the model's performance is expected to be somewhat affected. We anticipate some misclassifications involving CICY5, which could also impact the classification accuracy for CICY3 and CICY4. Nonetheless, these issues should not drastically affect the overall results. The performance outcomes of the classification models are summarized in Tables 4 and 5.

One observes that the naive Bayes algorithm performs exceptionally well in our multiclass classification task. The near-perfect learning outcome is a very positive result.

Finally, it is insightful to examine the naive Bayes density plots shown in Fig. 11, which illustrate the classification boundaries for our multiclass problem.

In these naive Bayes graphs, we observe overlapping densities for high values of $h^{1,1}$ and across most values of $h^{2,1}$. A possible explanation is that many of the Hodge numbers of CICY5 are still undetermined. However, it is evident that the densities remain well separated for $h^{1,1}$.

The support vector machine graph in the multiclass classification (Fig. 10(b)) reveals that parts of CICY4 are obscured by both CICY3 and CICY5. Interestingly, a distinct region in the Calabi–Yau three-fold landscape, characterized by $17 \leq h^{1,1} \leq 30$ and $20 \leq h^{2,1} \leq 40$, can be clearly identified. It is also notable that CICY5 overlaps with both CICY3 and CICY4 in the region $0 \leq h^{1,1} \leq 4$ and $20 \leq h^{2,1} \leq 40$, which is somewhat surprising.

This manuscript addresses the classification of Calabi–Yau manifolds using machine learning algorithms. Our investigations demonstrate that complete intersection Calabi–Yau three-folds, four-folds, and

[2] The SMOTE plus Tomek link method is not applicable here due to the very small number of CICY5 instances.

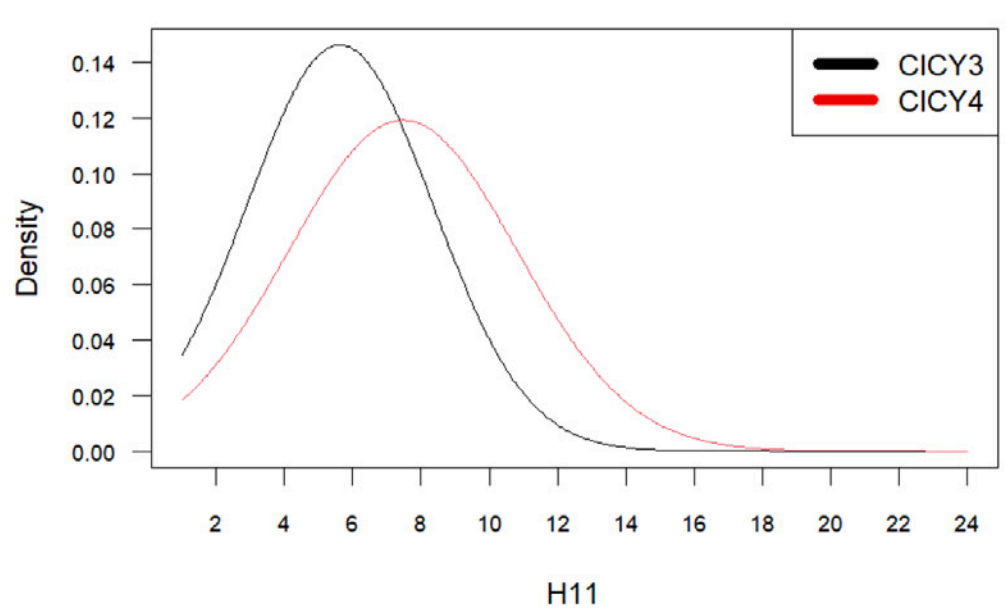

(a) Naive Bayes plot of H11.

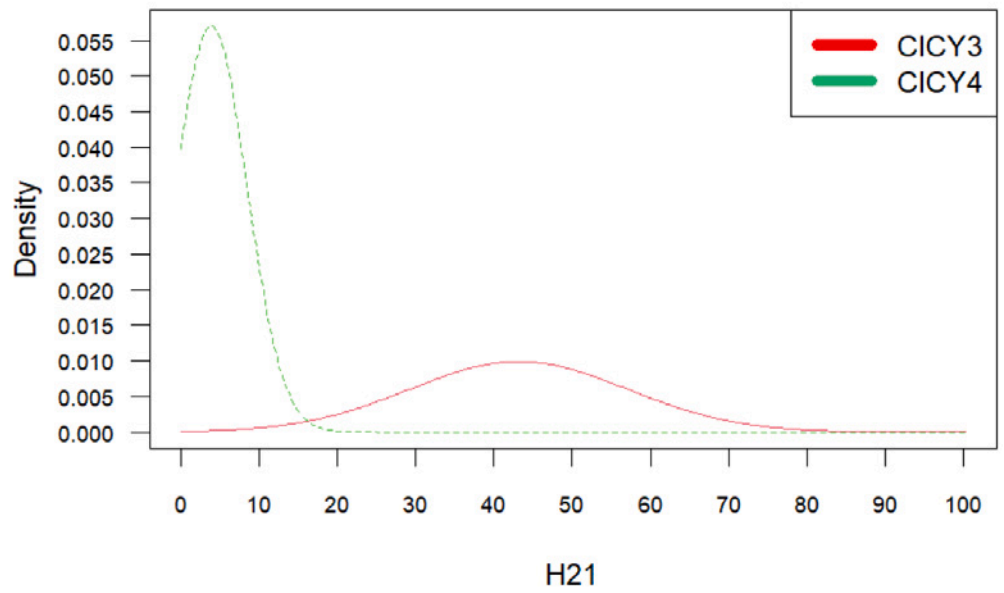

(b) Naive Bayes plot of H21.

**Fig. 9.** Naive Bayes plot of the Hodge numbers for CICY3 and CICY4.

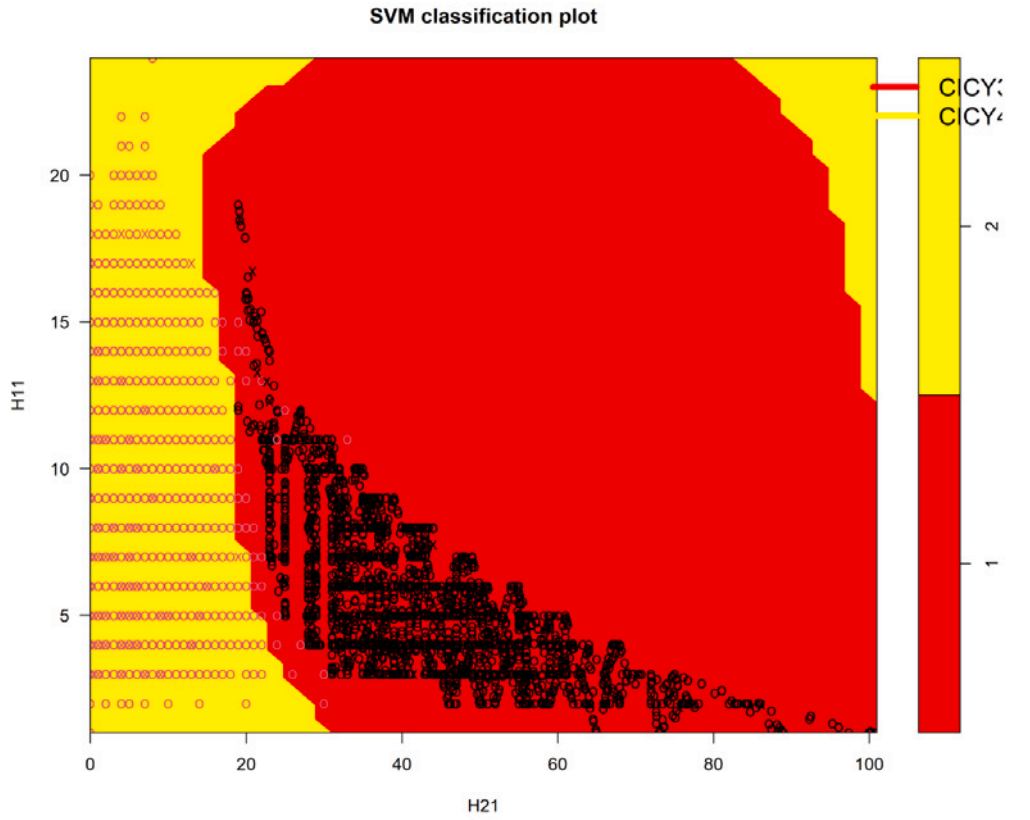

(a) SVM decision boundary plot of CICY3 and CICY4.

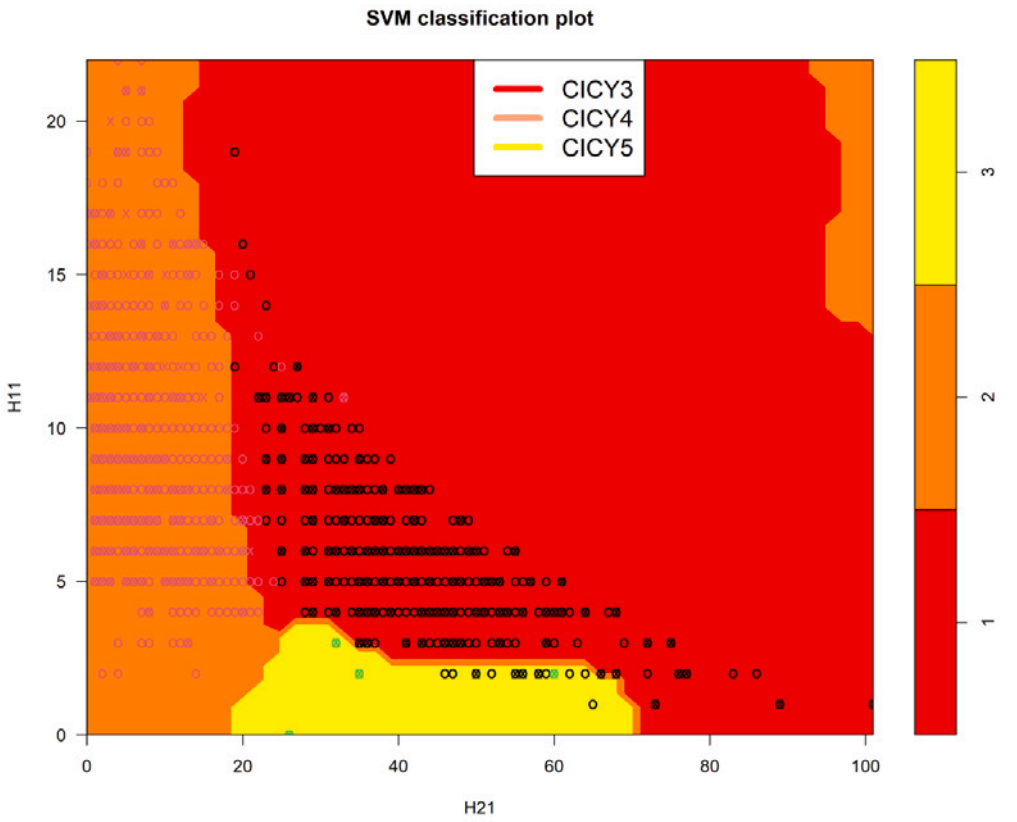

(b) SVM decision boundary plot of the three CICYs.

**Fig. 10.** SVM decision boundary plot of the Data from SVM classification.

**Table 4**
Parameters for some classification techniques (validation sets).

| Classification | Accuracy | F1-score | | | Confusion matrix | | | |
|---|---|---|---|---|---|---|---|---|
| Naives Bayes | 1.000000 | CICY3 | CICY4 | CICY5 | Class | CICY3 | CICY4 | CICY5 |
| | | 1.00000 | 1.00000 | 1.0000 | CICY3 | **590** | 0 | 0 |
| | | | | | CICY4 | 0 | **551** | 0 |
| | | | | | CICY5 | 0 | 0 | **562** |
| Ksvm | 0.9929536 | 0.9898 | 0.9982 | 0.9912 | CICY3 | **580** | 0 | 10 |
| | | | | | CICY4 | 2 | **549** | 0 |
| | | | | | CICY5 | 0 | 0 | **562** |
| Rpart | 0.9894304 | 0.9848 | 0.9890 | 0.9947 | CICY3 | **584** | 0 | 6 |
| | | | | | CICY4 | 12 | **529** | 0 |
| | | | | | CICY5 | 0 | 0 | **562** |
| Svm | 0.9770992 | 0.9659 | 0.9955 | 0.9706 | CICY3 | **553** | 3 | 34 |
| | | | | | CICY4 | 2 | **549** | 0 |
| | | | | | CICY5 | 0 | 0 | **562** |
| Lda | 0.8625954 | 0.8151 | 0.9679 | 0.8060 | CICY3 | **498** | 28 | 64 |
| | | | | | CICY4 | 0 | **543** | 8 |
| | | | | | CICY5 | 134 | 0 | **428** |

five-folds can be effectively classified by supervised machine learning techniques. For the binary classification of $CICY_3$ and $CICY_4$ using naïve Bayes, there is no overlap in the density plots, indicating that the two Hodge numbers $h^{1,1}$ and $h^{2,1}$ are highly discriminative. The densities are well separated, confirming that these Hodge numbers are very effective in distinguishing $CICY_3$ from $CICY_4$. Similarly, in the multiclass classification involving $CICY_3$, $CICY_4$, and $CICY_5$, the density plots show minimal overlap, reinforcing the suitability of Hodge numbers for classifying CICYs.

Nevertheless, the region of $CICY_4$ that remains unmasked by $CICY_3$ and $CICY_5$ warrants further detailed study. The results presented here provide a foundation for future work, such as applying association rule learning techniques to the classification of CICYs. The remaining codes implementing the machine learning algorithms and the associated packages used in this paper are available upon request.

**Table 5**
Parameters for some classifications techniques (training sets).

| Classification | Accuracy | F1-score | | | Confusion matrix | | | |
|---|---|---|---|---|---|---|---|---|
| | | CICY3 | CICY4 | CICY5 | Class | CICY3 | CICY4 | CICY5 |
| Naives Bayes | 0.931141 | 0.8879 | 0.9821 | 0.9209 | CICY3 | **1846** | 14 | 388 |
| | | | | | CICY4 | 64 | **2220** | 3 |
| | | | | | CICY5 | 0 | 0 | **2276** |
| Ksvm | 0.9913375 | 0.9868 | 0.9971 | 0.9900 | CICY3 | **2202** | 0 | 46 |
| | | | | | CICY4 | 13 | **2274** | 0 |
| | | | | | CICY5 | 0 | 0 | **2276** |
| Rpart | 0.9913375 | 0.9869 | 0.9910 | 0.9961 | CICY3 | **2230** | 0 | 18 |
| | | | | | CICY4 | 41 | **2246** | 0 |
| | | | | | CICY5 | 0 | 0 | **2276** |
| Svm | 0.8484151 | 0.84264 | 0.8785 | 0.7970 | CICY3 | **553** | 3 | 34 |
| | | | | | CICY4 | 18 | **3033** | 506 |
| | | | | | CICY5 | 13 | 288 | **1592** |
| Lda | 0.8284045 | 0.79439 | 0.8674 | 0.7610 | CICY3 | **170** | 27 | 0 |
| | | | | | CICY4 | 22 | **3043** | 492 |
| | | | | | CICY5 | 39 | 389 | **1465** |

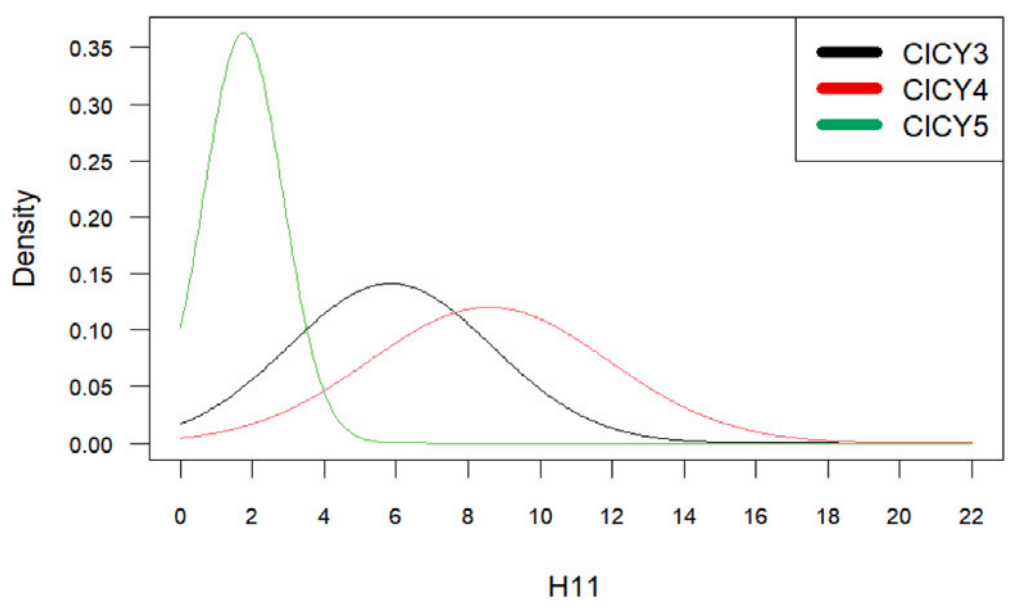

(a) Naive Bayes plots of H11.

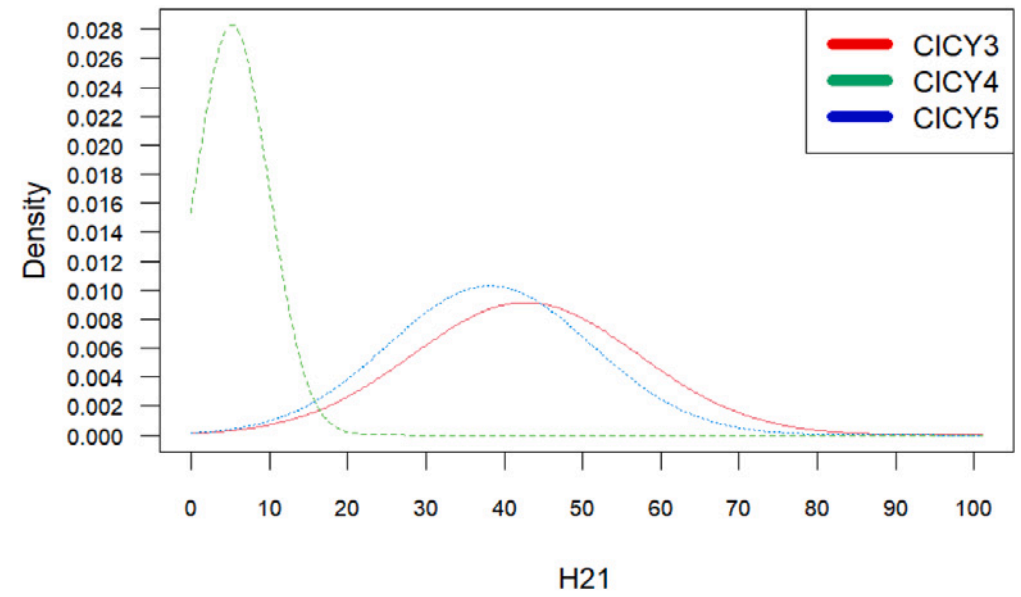

(b) Naive Bayes plots of H21.

**Fig. 11.** Naive Bayes plot of the Hodge numbers for CICY3, CICY4 and CICY5.

**Future Directions.** Several research avenues naturally follow from our findings. A closer study of the unmasked region of the $CICY_4$ landscape may reveal geometric features not captured by $(h^{1,1}, h^{2,1})$ alone. Addressing degeneracies in configuration matrices further motivates the use of symmetry-aware or equivariant architectures, while geometric or graph-based models capable of learning directly from configuration matrices offer another promising direction. In addition, association rule learning techniques – recently applied to higher-dimensional CICYs in [65] – provide an interpretable framework that could complement the present approach. Finally, extending our methodology to other Calabi–Yau families, such as generalized CICYs, may uncover broader structural patterns across the landscape.

These directions highlight how the present work lays a foundation for more comprehensive ML-based investigations of Calabi–Yau manifolds.

## CRediT authorship contribution statement

**Kaniba Mady Keita:** Writing – review & editing, Writing – original draft, Validation, Software, Project administration, Methodology, Investigation, Formal analysis, Conceptualization. **Younouss Hamèye Dicko:** Supervision.

## Declaration of competing interest

The authors declare that there is no conflict of Interest of whatsoever associate with this manuscript.

## Data availability

Data will be made available on request.